# Hybrid M-mode-like OCT imaging of 3D microvasculature in vivo using reference-free processing of complex-valued B-scans

Lev A. Matveev,[1,2,3,*] Vladimir Yu. Zaitsev,[1,2,3] Grigory V. Gelikonov,[1,2] Alexandr L. Matveyev,[1,2] Alexander A. Moiseev,[1] Sergey Ksenofontov,[1] Valentin M. Gelikonov,[1,2,3] Valentin Demidov,[4] Alex Vitkin,[2,4]

[1]Institute of Applied Physics RAS, 60390, Ulyanov St., 46, Nizhny Novgorod, Russia
[2]Nizhny Novgorod State Medical Academy, 603005, Minin and Pozharsky Sq., 10/1, Nizhniy Novgorod, Russia
[3]Nizhny Novgorod State University, 603950, Gagarina av., 23, Nizhny Novgorod, Russia
[4]University of Toronto and University Health Network, M5G 2M9, 610 University Ave., Toronto, Ontario, Canada
*Corresponding author: lionnn52rus@mail.ru

We propose a novel OCT-based method for visualizing microvasculature in 3D using reference-free processing of individual complex-valued B-scans with highly overlapped A-scans. In the lateral direction of such a B-scan, the amplitude and phase of speckles corresponding to vessel regions exhibit faster variability, and thus can be detected without comparison with other B-scans recorded in the same plane. This method combines elements of several existing OCT angiographic approaches, and exhibits: (i) enhanced robustness with respect to bulk tissue motion with frequencies up to tens of Hz; (ii) resolution of microcirculation images equal to that of structural images and (iii) possibility of quantifying the vessels in terms of their decorrelation rates.

http://dx.doi.org/10.1364/OL.99.099999

The need for increased information content in optical coherence tomography (OCT) of biological tissues has been exemplified by recent efforts to complement conventional structural images with additional contrast mechanisms, such as polarization-sensitive measurements and elastoraphic imaging [1-3], studies of rheological/relaxational characteristics [4,5], and microvasculature characterization [6-16], as summarized in recent reviews [17,18]. Many vasculature-imaging methods detect flow via phase-resolved Doppler effects, including either detailed quantification of flow profiles (e.g. [6]) or at least flow direction and approximate grading of flow velocities (power-Doppler approach [8,13]). In these methods, the Doppler frequency shift is estimated by comparing phases of adjacent A-scans. Closely related modified angiographic techniques employ additional modulation to ensure non-zero Doppler shift even from motionless tissue, followed by advanced filtering procedures to better single out moving scatterers from nearly motionless surrounding tissue [8,9,11].

Another group of angiographic methods (e.g., [12,14,15]) uses temporal variability of structural B-scan speckle patterns due to motion of scatterers in the liquid, comprising both collective flow and Brownian motion [14]. For mapping regions of increased speckle-texture variability, various analyses can be used, in particular correlation processing (e.g., correlation mapping, or Cm) [12] or speckle-variance (Sv) methods [14,15]. The common feature of these approaches is that they compare several consecutive B-scans obtained from the same location. However, the time of B-scan acquisition is typically 2-3 orders of magnitude longer than that of individual A-lines. For typical interval between B-scans of ~tens of milliseconds, several "repeated" B-scans should be compared for reliable distinction between the stable "solid" pixels and variable faster decorrelating "liquid" pixels; this methodology is prone to natural bulk motions of living tissue [15]. For example, in typical realizations of Sv approach, the decorrelation times of blood in vessels on the order $10^1$-$10^2$ ms require ~8 repeated B-scans in a stack [14,15], since typical B-scan frame rates range from several to several tens of Hz. Technological improvements enabling higher frame rates may decrease the bulk tissue motion artefacts, but may also decrease the Sv microvascular contrast due to insufficient decorrelation during the inter-frame interval. So the current compromise is to physically stabilize the interrogated tissue and/or attempt to eliminate the bulk-motion artifacts via post-processing. Another potential drawback of reported Sv and Cm inter B-scan analysis techniques is that all strongly decorrelated speckles in the resultant microvascular images look similar and do not retain information on decorrelation-time differences of different vessels.

Here, we propose a novel 3D angiographic approach which combines elements of Doppler-based methods and Sv methods based on amplitude and phase variability of speckles corresponding to moving scatterers in blood compared to a more "stable" background bulk tissue. It represents reference-free processing of individual complex-signal B-scans in which the horizontal step between the adjacent A-scans is significantly smaller than the optical beam diameter. In view of such features, we refer to this as a "M-mode-like speckle variability" (MMLSV) approach. The contrast mechanism here is similar to previous Sv/Cm methods but signal acquisition / analysis are different – in image regions corresponding to the blood vessels, the complex field experiences higher spatio-temporal variability (comprising both its amplitude and phase), whereas for fragments of densely spaced A-scans corresponding to "solid" tissue, the structure of complex-field pixels is almost identical. The proposed MMLSV method thus combines the contrast mechanisms used in Sv/Cm and Doppler OCT, which is favorable for its sensitivity and robustness. However, the single densely-sampled complex B-scan processing in MMLSV is significantly distinct from the previous methods. There, the variability of the complex field within vessels is detected by direct comparison of amplitudes/phases of either corresponding individual pixels or small groups of pixels in consecutively obtained B-scans in the same plane (Sv and Cm methods), or in preselected pairs of closely located A-scans in a B-scan (Doppler methods). Here, MMLSV analyzes the variability of the complex field in each single dense B-scan by evaluating (in an integral sense) groups of closely located A-scans via high-pass filtering of horizontal spatial Fourier components. This filtering makes it possible to preferentially segment out regions of increased complex-field variability corresponding to flow and/or Brownian motion of blood scatterers in the microvasculature. Further, by varying the threshold frequency of this high-pass filtering, the regions with different degree of complex-speckle variability can be detected, making it possible to grade faster/slower decorrelating vessels. Note that the high-pass filtering does not single out the Doppler optical components in the signal *per se*, but rather estimates the amplitude-phase variability of speckles along the horizontal coordinate in oversampled B-scans.

The essence of the proposed MMLSV approach is illustrated in Fig.1, where panel (a) schematically shows a stack of B-scans with strongly overlapped A-scans filling the inspected 3D volume. The experimental B-scan on the top of panel (d) illustrates that indeed in such dense B-scans, the motionless scatterers in the structural image look as horizontally elongated speckles (shown as "long dashes" in the corresponding schematic below). The encircled region in this B-scan fragment shows a vessel cross section with moving scatterers (erythrocytes and other blood cells). This motion causes faster varying speckles that are much shorter in the horizontal direction along B-scan ("short-dash" speckles in the schematic below). The horizontal spatial spectrum of the shorter dashes thus extends to higher spatial frequencies. Consequently, applying high-pass filtering to such strongly oversampled B-scans predominantly retains higher variability "short-dash" speckles representing moving scatterers. Panel (c) shows that by changing the threshold frequency of the high-pass filter, areas with faster and slower varying pixels can be distinguished; this leads to vessel gradation / flow quantification as discussed below (Figure 2). After performing inverse Fourier transform, the so-filtered B-scan exhibits only the regions corresponding to blood-containing faster varying speckles (panel (e), filtered experimental image on top and schematic on bottom). The entire stack of processed B-scans (panel (g)) enables 3D visualization of the microvasculature. Resultant 3D images can be color-encoded to represent vessels with higher/lower temporal variability (analogous to band-pass Doppler OCT [7]), or to encode depth (similar to other vasculature-imaging methods [14,15]). The depth-encoded result is displayed in panel (f), along with a photograph of the dorsal window chamber in a mouse showing the location of the imaged volume. Once again, note that the difference between high and low frequency components is not related to the Doppler shift of optical frequencies, but rather indicates different rate of variability of speckles. For the motionless scatterers, this is determined by the time required for the scanning optical beam to cover its own diameter, and the motion of scatterers further reduces this characteristic time (much like in Sv methods).

Another important and inherently advantageous step of image processing in MMLSV is illustrated in panel Fig. 1b, allowing for rather efficient compensation of natural bulk tissue motions. Indeed, due to relatively small interval between A-scans, the axial bulk tissue motion displacements ("clutter") can remain smaller than the quarter of optical wavelength for bulk-motion velocities up to ~several cm/s. Determining the phase difference $[\varphi_{n+1} - \varphi_n]_{aver}$ for the neighboring A-scans numbered $n$ and $n+1$ (e.g., using the Kasai estimator [19]), the averaged phase differences can be compensated, so that the equalized phase $\varphi_{n+1}(i)_{eq}$ of $i_{th}$ pixel in $(n+1)_{th}$ A-scan becomes

$$\varphi_{n+1}(i)_{eq} = \varphi_{n+1}(i) - [\varphi_{n+1} - \varphi_n]_{aver}$$

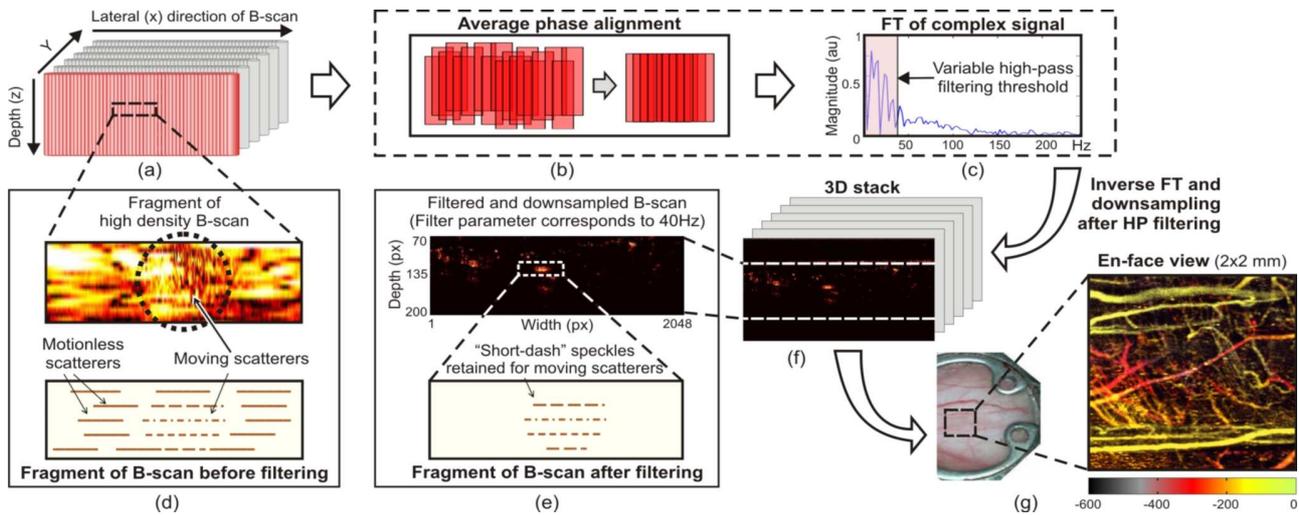

Fig. 1. Schematic elucidation of the NMMSv approach: (a) – A stack of dense B-scans with strongly overlapped A-scans; (b) - a single B-scan before and after equalization of average phases of neighboring A-scans; (c) – spatial spectrum of a single horizontal array of pixels in a dense B-scan, where low-frequency components correspond to slowly varying motionless scatterers. (d) – experimnetal B-scan fragment with encircled region of rapidly varying speckles corresponding to a vessel cross-section (top) and a schematic B-scan with speckles slowly and rapidly varying in the horizontal direction (bottom); (e) – same as (d), but after high-pass filtering (only vessels cross-sections are now visible); (f) shows an photo of a mouse dorsal window chamber and an example of *en-face* depth-encoded view of a 2mm x 2mm (lateral) 0-0.6 mm (depth) region of the detected mouse microvasculature.

Such equalization (see also [20, 21]) ensures significantly higher stability of the complex-field signal produced by the scatterers in the "solid" tissue. It should be emphasized that for the small portions of A-scans corresponding to blood vessel cross sections, such phase equalization over entire A-scans does not cancel the variations of the complex signal from scatterers moving in the liquid, so that the masking influence of the axial bulk motion clutter motions can be effectively reduced. Concerning the tolerance to clutter-motion components in the horizontal directions, the signal from "solid" pixels remains fairly stable even if horizontal displacements are comparable with the optical-beam diameter during the interval of analysis (~the number of nearly overlapping A-scans divided by A-scan rate). Although this interval in MMLSV should be significantly greater than the interval between individual A-scans, the allowable horizontal displacements can be an order of magnitude greater than the above-mentioned axial ones. Thus, for both vertical and horizontal clutter motions, the proposed processing is approximately equally tolerant up to velocities of order ~cm/s. Comparison of several consecutive B-scans via either speckle variance [14,15] or correlational mapping [7] methods requires significantly higher stability (because the allowable displacements are of the same order, but the interval between processed B-scans is significantly greater).

The above-described methodology generates images that are initially highly oversampled in the horizontal plane. For convenience of visualization, they can be down-sampled to the natural horizontal resolution of the OCT scanner, as determined by the optical beam diameter. Thus another advantage of MMLSV approach is that the resolution of resultant microvascular images is the same as the OCT system's structural images (for example, not compromised by the correlation-window size as in Cm methods).

For experimental demonstrations of the proposed NNMSV methodology, we used a home-built Fourier domain spectral OCT scanner with the central optical wavelength $\lambda = 1.32$ μm and a bandwidth of 106 nm and a rate of 20 kHz for spectral fringes (yielding 10 kHz rate of the formed and visualized complex-valued A-scans). The axial and lateral resolutions of the system are 10 μm and 20 μm, respectively. In the depth direction, the spectrometer array enables 256 pixels and the chosen number of B-scans for 3D scanning also equals 256. For these system parameters, phase equalization (Figure 1(b)) ensures rather efficient compensation of bulk tissue motions with characteristic frequencies ~10-20 Hz for displacement amplitudes ~0.2-0.1 mm.

For the examples illustrated below, the density of complex A-scans within a B-mode scan is 16,384 A-scan/mm. For this regime, the *en face* area of 2x2 mm$^2$ required ~13 min scanning time, which is quite long but can be additionally optimized. The number of overlapped A-scans divided by their rate determines the maximal intervals for observing speckle variability; for the discussed regime, this is ~30 ms. Thus, the observable characteristic times of speckle variability related to intrinsic motions in "liquid" pixels should be smaller than this interval. Evidently, denser overlapping increases sensitivity to slowly decorrelating vessels, but increases the scanning time.

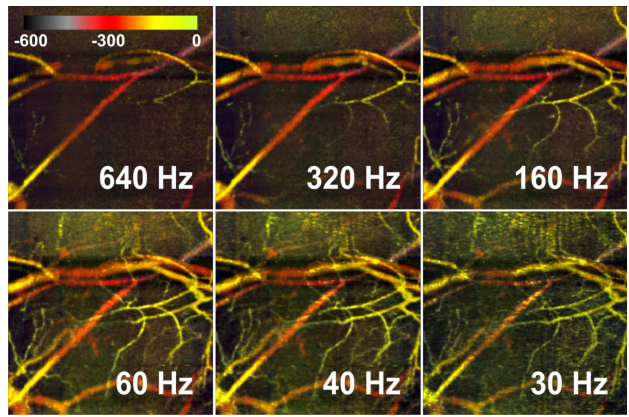

Fig. 2. Examples of color depth encoded *en-face* microvasculature images obtained for different filter thresholds $F_{min}$, demonstrating gradual appearance of finer vessels with different rates of speckle variability. Presumably, for the highest $F_{min} > 160$ Hz, the contributions of faster flows are dominant (first arterioles and then venules), whereas for $F_{min}$~40-60 Hz, the slower regions of flow as well as possibly Brownian scatterer motion in finer microvessels (~capillaries) contribute to the angiographic image as well. The signal from the motionless background tissue begins to degrade the image quality at $F_{min} = 30$ Hz. FOV = 2mm x 2mm for all panels. Another example of photographic image superposed with a so-obtained angiogram is given in Supplement_2.

Figure 2 shows some examples of *en-face* depth-encoded images obtained by applying different threshold frequencies for high-pass filtering of dense B-scans, followed by down-sampling to the natural lateral resolution of the OCT scanner. All animal procedures were performed in accordance with approved Animal Use Protocol at the Nizhny Novgorod State Medical Academy. In these experiments, we used BalbC mice with implanted dorsal window chambers (similar to [14,15]). To emphasize the method's robustness, we note that because of small gap beneath the window glass, mouse's skin exhibited significant natural motions both in axial and lateral directions (see movie in Supplement_1), yet successful imaging was possible.

Note that the finite number of overlapping A-scans means that even speckles corresponding to motionless scatterers exhibit variability along B-scan, which for the examples in Fig. 2 corresponds to the ultimately low temporal variability rate $F_{ult} \sim 30Hz$. By this reason for filtering threshold $F_{min} = 30Hz \sim F_{ult}$, bulk tissue begins to degrade the angiographic image. Choosing $F_{min} > F_{ult}$ predominantly retains image regions with characteristic variability intervals (decorrelation times) smaller than $1/F_{min}$. Figure 2 shows that by decreasing $F_{min}$ towards $F_{ult}$, slower deccorrelating vessels can be visualized (although the noise from motionless/slower moving background gradually increases). The rate of this decorrelation is determined by both regular (convective) flow in a vessel and Brownian scatterer (erythrocyte) motion (that can actually dominate in Sv imaging [14]). The latter fact impedes straightforward usage of simplified arguments [22] based on the assumption of scatterer motion like in moving solid body to link the flow and speckle variability and.

In this context, one can point out two potentially relevant characteristic times: (1) ~ $D/V_{sc}$, for a scatterer moving with velocity $V_{sc}$ crossing the beam diameter $D$, and (2) ~ $\lambda/4V_{sc}^{rel}$ during which the axial distance between a pair of subresolution scatterers having the relative axial velocity $V_{sc}^{rel}$ varies by a quarter of wavelength $\lambda/4$. Such variation can also cause speckle blinking in the image [18]. For parameters used in obtaining Fig. 2, characteristic $V_{sc}$ and $V_{sc}^{rel}$ differ ~20X and depending on $F_{min}$ fall into the range from fractions mm/s to several mm/s, which is realistic for Brownian motion and collective (convective) flow velocities.

Overall, the proposed MMLSV approach represents a hybrid method exhibiting certain features of Doppler and phase-resolved methods that compare pixel phases in individual A-scans (or use other ways to estimate Doppler frequency components, for example filtering). From the viewpoint of scanning regime, the MMLSV is close to M-mode, whereas the use of speckle variability resembles elements of Sv/Cm methods, although here the comparison is made within each dense B-scan rather than between repeated B-scans from the same position. The use of both amplitude and phase information MMLSV-OCT further enhances the positive traits of the above-mentioned approaches. Note that the utility of combining analogous techniques is also shown in recent work [23] using a two-stage combined processing. The equalization of averaged phases of neighboring A-scans significantly reduces the noise influence of clutter-type bulk tissue motion phase variations, in comparison with conventional Doppler and with Sv/Cm OCT methods. Quantification of blood vessels in terms of faster/slower variability is also naturally enabled thru the user-selected cut-off frequency of the high-pass filtering, and may be linked to physiologically important dynamic variables such as flow velocity and perfusion. This, and the separation of flow versus Brownian motion contributions to MMLSV signal represent interesting challenges that will be addressed in future publications.

The authors thank the researchers of the OCT Bio-imaging Laboratory at the Nizhny Novgorod State Medical Academy for help with the animal experiments. This work was supported by the grant of the Russian Federation Government No 14.B25.31.0015 and RFBR grant No 13-02-97131. VYZ and VMG acknowledge partial support by the contract No 02.B49.21.0003 between the Russian Ministry of Education and Nizhny Novgorod State University.